# Prodromal Diagnosis of Lewy Body Diseases Based on the Assessment of Graphomotor and Handwriting Difficulties

Zoltan Galaz[1], Jiri Mekyska[1], Jan Mucha[1], Vojtech Zvoncak[1], Zdenek Smekal[1], Marcos Faundez-Zanuy[2], Lubos Brabenec[3], Ivona Moravkova[3,4,5], and Irena Rektorova[3,4]

[1] Department of Telecommunications, Faculty of Electrical Engineering and Communication, Brno University of Technology, Brno, Czech Republic
xgalaz00@gmail.com
[2] Escola Superior Politecnica, Tecnocampus, Mataro, Barcelona, Spain
[3] Applied Neuroscience Research Group, Central European Institute of Technology – CEITEC, Masaryk University, Brno, Czech Republic
[4] First Department of Neurology, Faculty of Medicine and St. Anne's University Hospital, Masaryk University, Brno, Czech Republic
[5] Faculty of Medicine, Masaryk University, Brno, Czech Republic

**Abstract.** To this date, studies focusing on the prodromal diagnosis of Lewy body diseases (LBDs) based on quantitative analysis of graphomotor and handwriting difficulties are missing. In this work, we enrolled 18 subjects diagnosed with possible or probable mild cognitive impairment with Lewy bodies (MCI-LB), 7 subjects having more than 50% probability of developing Parkinson's disease (PD), 21 subjects with both possible/probable MCI-LB and probability of PD > 50%, and 37 age- and gender-matched healthy controls (HC). Each participant performed three tasks: Archimedean spiral drawing (to quantify graphomotor difficulties), sentence writing task (to quantify handwriting difficulties), and pentagon copying test (to quantify cognitive decline). Next, we parameterized the acquired data by various temporal, kinematic, dynamic, spatial, and task-specific features. And finally, we trained classification models for each task separately as well as a model for their combination to estimate the predictive power of the features for the identification of LBDs. Using this approach we were able to identify prodromal LBDs with 74% accuracy and showed the promising potential of computerized objective and non-invasive diagnosis of LBDs based on the assessment of graphomotor and handwriting difficulties.

This work was supported by grant no. NU20-04-00294 (Diagnostics of Lewy body diseases in prodromal stage based on multimodal data analysis) of the Czech Ministry of Health and by Spanish grant of the Ministerio de Ciencia e Innovación no. PID2020-113242RB-I00.

## 1 Introduction

Lewy body diseases (LBDs) is a term describing a group of neurodegenerative disorders characterized by a pathophysiological process of $\alpha$-synuclein accumulation in specific brain regions leading to the formation of Lewy bodies and Lewy neurites resulting in cell death. LBDs consists of two major clinical entities: Parkinson's disease (PD) and dementia with Lewy bodies (DLB) [29,38]. Although the phenotypes and temporal evolution of motor and cognitive symptoms of these two diseases vary, they share many clinical and pathophysiological features and are therefore referred to as LBDs spectrum. Together with Alzheimer's disease (AD), LBDs comprise the major part of all cases of neurodegenerative disorders.

It is known that LBDs do not start suddenly. At the time the clinical symptoms occur, the neurodegenerative process has reached a severe degree in which most of the targeted neurons have already been damaged. Before the clinical diagnosis based on the presence of typical clinical symptoms becomes possible, there is a long period of the underlying neurodegenerative process with subtle or nonspecific symptoms [18,29] such as sleep disturbances, mood changes, smell loss, constipation, etc. This period of LBDs is called the prodromal stage.

One of the early markers of PD is PD dysgraphia (micrographia and other alterations in handwriting, e.g. kinematic and dynamic) [21,32,33]. Similarly, some manifestations of dysgraphia have been observed in the prodromal DLB as well [23]. Although modern approaches to the analysis of graphomotor and handwriting difficulties (utilising digitising tablets) were proved to work well during e.g. diagnosis of the clinical stage of PD [9,11,35], assessment of cognition in PD patients [4], or discrimination of AD and mild cognitive impairment (MCI) [15], to the best of our knowledge, no studies employed this technology (with high potential) in the prodromal diagnosis of LBDs in a larger scale.

Identification of the early stages of LBDs is crucial for the development of disease-modifying treatment since the neurodegeneration may be possibly stopped or treated before the pathological cascades start. Therefore, the goal of this study is to explore whether the computerised assessment of graphomotor and handwriting difficulties could support the prodromal diagnosis of LBDs, more specifically, we aim to:

1. identify which task significantly discriminates LBD patients and age- and gender-matched healthy controls (HC),
2. identify what conventional online handwriting features have good discrimination power.

## 2 Materials and Methods

### 2.1 Dataset

We enrolled 39 subjects (19 females, 20 males, age = 69.53± 6.61) diagnosed with possible or probable MCI (based on the scores of the MoCA – Montreal Cognitive Assessment [25] and based on the CCB – Complex Cognitive Battery, see the explanation below) who were simultaneously diagnosed with possible or probable MCI-LB (i.e. mild cognitive impairment with Lewy bodies) based on the criteria published by McKeith et al. [22]. In this group, 21 subjects also had more than 50% probability of developing PD (calculated following the MDS criteria published in [18]). In addition, we enrolled 7 subjects (2 females, 5 males, age = 66.41± 4.32) without possible/probable MCI-LB, but still with more than 50% probability of developing PD. Finally, we enrolled 37 HC (26 females, 11 males, age = 67.60 ± 5.61). In the experiments, we stratified the subjects into two groups, HC vs. LBD (i.e. people with a high risk of developing PD or DLB).

CCB was used to evaluate four cognitive domains: 1) memory (The Brief Visuospatial memory test–revised [2], Philadelphia Verbal Learning Test [3]); 2) attention (Wechsler Adult Intelligence Scale-III: Letter-Number Sequencing, Digit Symbol Substitution [37]); 3) executive functions (Semantic and phonemic verbal fluency [30], Picture arrangement test [37]); and 4) visuospatial functions (Judgment of Line Orientation [36]). The cognitive domain z-scores were computed as the average z-scores of the tests included in the particular domain.

The participants were asked to perform a set of three tasks:

1. Archimedean spiral (spiral) – we consider this task as a graphomotor one, i.e. it is a building block of some letter shapes; in addition, it is a golden standard in PD dysgraphia diagnosis [35]
2. sentence "Tramvaj dnes už nepojede" (translation: "A tram will not go today.") writing (sentence) – this handwriting task was used e.g. in the PaHaW database [11]
3. pentagon copying test (pentagons) – it is a task frequently used for quantification of cognitive decline [4]

All participants were right-handed and had Czech as their native language. They all signed an informed consent form that was approved by the local ethics committee.

### 2.2 Feature Extraction

The participants were asked to perform the tasks (using the Wacom Ink pen) on an A4 paper that was laid down and fixed to a digitizing tablet Wacom Intuos 4 M (sampling frequency $f_s$ = 130 Hz). Before the acquisition, they had some time to get familiar with the hardware. The recorded time series (x and y position; timestamp; a binary variable, being 0 for in-air movement and 1 for on-surface movement, respectively; pressure exert on the tablet's surface during writing; pen tilt; azimuth) were consequently parameterised utilising the following set of features (we selected the set based on available reviews and based on our experience [9,11,35]):

1. temporal – duration of writing, ratio of the on-surface/in-air duration, duration of strokes, and ratio of the on-surface/in-air stroke duration
2. kinematic – velocity, and acceleration
3. dynamic – pressure, tilt, and azimuth
4. spatial – width, height, and length of the whole product, as well as its particular strokes, i.e. stroke width, height, and length
5. spiral-specific – degree of spiral drawing severity [31], mean drawing speed of spiral [31], second-order smoothness of spiral [31], spiral precision index [5], spiral tightness [31], variability of spiral width [31], and first-order zero-crossing rate of spiral [31]
6. other – number of interruptions (pen elevations), number of pen stops [27], tempo (number of strokes normalised by duration), number of on-surface intra-stroke intersections, relative number of on-surface intra-stroke intersections, number of on-surface inter-stroke intersections, and relative number of on-surface inter-stroke intersections, Shannon entropy [4], number of changes in the velocity profile, relative number of changes in the velocity profile

Most of the features were extracted using the recently released Python library handwriting-features (v 1.0.1) [14], the rest of them were coded in Matlab. Some features (mainly spatial, temporal and kinematic) were extracted from both on-surface and in-air movements. In addition, kinematic features were also analysed in horizontal and vertical projection. Features represented by vectors were consequently transformed to a scalar value using median, non-parametric coefficient of variation (nCV; interquartile range of feature divided by its median), slope and 95th percentile (95p).

## 2.3 Statistical Analysis and Machine Learning

To compare the distribution of features between the HC and LBD subjects, we conducted Mann-Whitney U-test with the significance level of 0.05. Moreover, to assess the strength of a relationship between the features and the subject's clinical status (HC/LBD), we computed Spearman's correlation coefficient ($\rho$) with the significance level of 0.05. Finally, during this exploratory step, we calculated Spearman's correlation with the domains of CCB and the overall score of MDS–Unified Parkinson's Disease Rating Scale (MDS–UPDRS), part III (motor part) [16].

To identify the presence of graphomotor or handwriting difficulties, we built binary classification models using an ensemble extreme gradient boosting algorithm known as XGBoost [6] (with 100 estimators). This algorithm was chosen due to its robustness to outliers, ability to find complex interactions among features as well as the possibility of ranking their importance. To build models with an optimal set of hyperparameters, we conducted 1000 iteration of randomized search strategy via stratified 5-fold cross-validation with 10 repetitions aiming to optimize balanced accuracy score (BACC; described in more detail along with other evaluation scores below). The following set of hyperparameters were optimized: the learning rate [0.001, 0.01, 0.1, 0.2, 0.3], $\gamma$ [0, 0.05, 0.10, 0.15, 0.20, 0.25, 0.5], the maximum tree depth [6, 8, 10, 12, 15], the fraction of observations to be randomly sampled for each tree (subsample ratio) [0.5, 0.6, 0.7, 0.8, 0.9, 1.0], the subsample ratio for the columns at each level [0.4, 0.5, 0.6, 0.7, 0.8, 0.9, 1.0], the subsample ratio for the columns when constructing each tree [0.4, 0.5, 0.6, 0.7, 0.8, 0.9, 1.0], the minimum sum of the weights of all observations required in a child node [0.5, 1.0, 3.0, 5.0, 7.0, 10.0], and the balance between positive and negative weights [1, 2, 3, 4].

The classification test performance was determined using the following classification metrics: Matthew's correlation coefficient (MCC), balanced accuracy (BACC), sensitivity (SEN) also known as recall (REC), specificity (SPE), precision (PRE) and F1 score (F1). These metrics are defined as follows:

$$\text{MCC} = \frac{TP \times TN + FP \times FN}{\sqrt{N}}, \quad (1)$$

$$\text{BACC} = \frac{1}{2}\left(\frac{TP}{TP + FN} + \frac{TN}{TN + FP}\right), \quad (2)$$

$$\text{SPE} = \frac{TN}{TN + FP}, \quad (3)$$

$$\text{PRE} = \frac{TP}{TP + FP}, \quad (4)$$

$$\text{REC} = \frac{TP}{TP + FN}, \quad (5)$$

$$\text{F1} = 2\frac{PRE \times REC}{PRE + REC} \quad (6)$$

where $N = (TP + FP) \times (TP + FN) \times (TN + FP) \times (TN + FN)$, $TP$ (true positive) and $FP$ (false positive) represent the number of correctly identified LBD subjects and the number of subjects incorrectly identified as having LBDs, respectively. Similarly, $TN$ (true negative) and $FN$ (false negative) represent the number of correctly identified HC and the number of subjects with LBDs incorrectly identified as being healthy.

To further optimize the trained classification models, we fine-tuned the models' decision thresholds via the receiver operating characteristics (ROC) curve. Using the fine-tuned decision thresholds, we evaluated the classification performance of the models using the leave-one-out cross-validation. The ROC curves were plotted using the probabilities of the predicted labels obtained via the cross-validation procedure that was employed during the final evaluation of the fine-tuned models.

And finally, to evaluate the statistical significance of the prediction performance obtained by the built classification models, a non-parametric statistical method named permutation test was employed [7,28]. For this purpose, we applied 1 000 permutations with the significance level of 0.05. To estimate the performance of the models on the permuted data, we used the same classification setup as employed during the training phase [26].

## 3 Results

The results of the exploratory data analysis are summarized in Table 1 (sorted based on the p-value for the Mann-Whitney U-test). The following features were found as the most distinguishing ones in terms of the differentiation between HC and subjects with LBD (the top 4 features are listed; *, **, and *** denote the p-values for both the Mann-Whitney U-test and Spearman's correlation coefficient being bellow the significance level of 0.05, 0.01, and 0.001, respectively; if both p-values are bellow a different significance level, the weaker statistical significance is selected): a) spiral – nCV of acceleration (on-surface) $\rho = -0.2438^*$, variability of spiral width $\rho = 0.2439^*$, median of azimuth $\rho = 0.2378^*$, and spiral precision index $\rho = 0.2367^*$; b) sentence – number of pen stops $\rho = 0.3460^{**}$, slope of duration of stroke (in-air) $\rho = 0.2823^{**}$, median of vertical velocity (on-surface) $\rho = -0.2438^*$, and median of vertical acceleration (on-surface) $\rho = 0.2317^*$; and c) pentagons – width of writing (on-surface) $\rho = -0.3045^{**}$, median of length of stroke (on-surface) $\rho = -0.2894^{**}$, nCV of length of stroke (on-surface) $\rho = 0.2489^*$, and median of duration of stroke (on-surface) $\rho = -0.2327^*$.

**Table 1.** Results of the exploratory analysis.

| Feature | p(U) | $\rho$ | p($\rho$) |
|---|---|---|---|
| Spiral | | | |
| nCV of acceleration (s) | 0.0138 | −0.2438 | 0.0263 |
| Variability of spiral width | 0.0138 | 0.2439 | 0.0263 |
| Median of azimuth | 0.0158 | 0.2378 | 0.0304 |
| Spiral precision index | 0.0162 | 0.2367 | 0.0312 |
| nCV of duration of stroke (s) | 0.0438 | −0.1892 | 0.0867 |
| Sentence | | | |
| Number of pen stops | 0.0009 | 0.3460 | 0.0014 |
| Slope of duration of stroke (a) | 0.0054 | 0.2823 | 0.0097 |
| Median of vertical velocity (s) | 0.0138 | −0.2438 | 0.0263 |
| Median of vertical acceleration (s) | 0.0182 | 0.2317 | 0.0351 |
| Rel. total number of intra-stroke intersections | 0.0232 | −0.2206 | 0.0451 |
| Pentagons | | | |
| Width of writing (s) | 0.0030 | −0.3045 | 0.0051 |
| Median of length of stroke (s) | 0.0045 | −0.2894 | 0.0080 |
| nCV of length of stroke (s) | 0.0123 | 0.2489 | 0.0233 |
| Median of duration of stroke (s) | 0.0178 | −0.2327 | 0.0343 |
| Median of horizontal acceleration (s) | 0.0182 | 0.2317 | 0.0351 |

p(U) – p-value of Mann-Whitney U-test; $\rho$ – Spearman's correlation coefficient; p($\rho$)– p-value of $\rho$; (s) – on-surface movement; (a) – in-air movement.

Next, Table 2 presents the results of the correlation analysis (*, and ** denote the p-values for Spearman's correlation coefficient being below the significance level of 0.05 and 0.01, respectively) between the features summarized in Table 1 and the following clinical information: a) MDS–UPDRS, and b) CCB domains.

**Table 2.** Results of the correlation analysis.

| Feature | $\rho$ (UPDRS) | $\rho$ (V) | $\rho$ (A) | $\rho$ (E) |
|---|---|---|---|---|
| Spiral | | | | |
| nCV of acceleration (s) | −0.3411* | −0.0013 | 0.1130 | 0.1899 |
| Variability of spiral width | 0.1653 | −0.3973** | −0.2981* | −0.1666 |
| Median of azimuth | 0.0442 | −0.3656* | −0.1029 | −0.0490 |
| Spiral precision index | 0.0606 | −0.0942 | −0.3987** | −0.2126 |
| nCV of duration of stroke (s) | −0.1089 | −0.1344 | −0.1618 | −0.0469 |
| Sentence | | | | |
| Num. of pen stops | −0.1018 | −0.1181 | 0.1012 | −0.1956 |
| Slope of duration of stroke (a) | 0.2620 | −0.1928 | −0.0513 | −0.1025 |
| Median of vertical velocity (s) | 0.0314 | 0.1106 | 0.0025 | 0.1794 |
| Median of vertical acceleration (s) | −0.2641 | −0.0301 | 0.3246* | 0.0193 |
| Rel. total num. of intra-stroke intersections | 0.0477 | 0.1647 | 0.1143 | 0.0962 |
| Pentagons | | | | |
| Width of writing (s) | −0.3448* | 0.2947* | 0.1351 | 0.1362 |
| Median of length of stroke (s) | −0.1545 | 0.1607 | 0.0501 | 0.1511 |
| nCV of length of stroke (s) | 0.3065* | −0.2435 | −0.1126 | −0.1155 |
| Median of duration of stroke (s) | −0.0348 | 0.0080 | −0.0085 | −0.0269 |
| Median of horizontal acceleration (s) | 0.3215* | −0.0226 | −0.1632 | −0.2060 |

$\rho$ – Spearman's correlation coefficient (∗ denotes p-value < 0.05 and ∗∗ denotes p-value < 0.01); UPDRS – MDS–Unified Parkinson's Disease Rating Scale, part III (motor part) [16]; V – visuospatial domain of CCB; A – attention domain of CCB; E – executive functions domain of CCB; (s) – on-surface movement; (a) – in-air movement.

To visualize the difference in the distribution of the top 4 features summarized above for HC and subjects with LBD, the box-violin plots are presented in Figs. 1, 2 and 3. The Fig. 1 shows the distribution of the features for the spiral drawing, the Fig. 2 shows the distribution of the features for the sentence writing, and the Fig. 3 is dedicated to the distribution of the features for the pentagon copying test.

The results of the classification analysis are summarized in Table 3. We trained 4 models in total: 3 models dedicated to each task separately and a model combining all of the tasks. The following results were achieved (where * and ** denote p-value of the permutation test bellow < 0.05 and < 0.01, respectively): a) spiral – BACC = 0.6848**, SEN = 0.8696, SPE = 0.5000; b) sentence – BACC = 0.7283**, SEN = 0.9783, SPE = 0.4783 c) pentagons – BACC = 0.6848**, SEN = 0.9348, SPE = 0.4348; and d) all tasks combined –

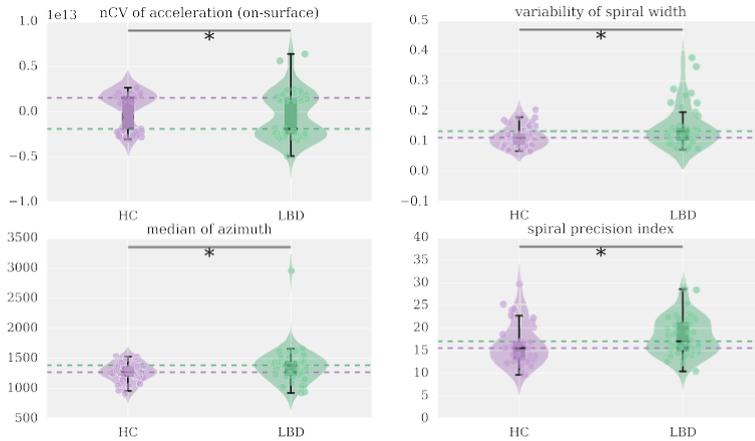

**Fig. 1.** Distribution of the top 4 most discriminating features (spiral drawing).

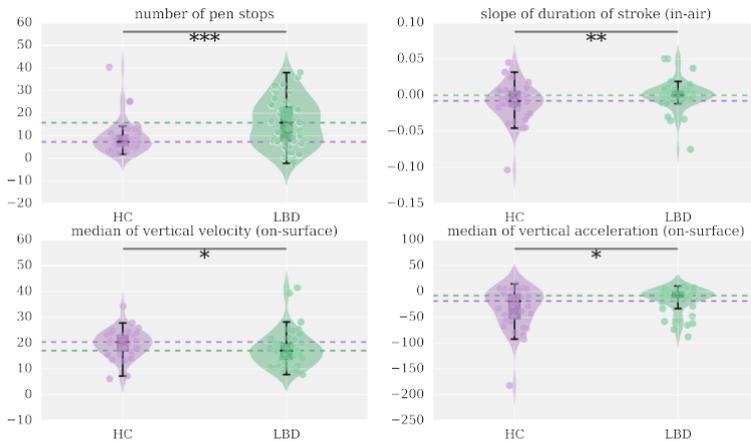

**Fig. 2.** Distribution of the top 4 most discriminating features (sentence writing).

BACC = 0.7391**, SEN = 0.8043, SPE = 0.6739. The ROC curves of the trained models are shown in Fig. 4.

## 4 Discussion

As mentioned in the methodology, the Archimedean spiral is considered as a gold standard, especially in the assessment of graphomotor difficulties in PD patients [5,8,31], nevertheless, it has been utilised during the quantitative analysis of Huntington's disease, essential tremor, or brachial dystonia as well [13]. Concerning the spiral features with the highest discrimination power (as identified by the Mann-Whitney U-test), we observed that the LBD group was associated with a lower range in on-surface acceleration, which we suppose is caused

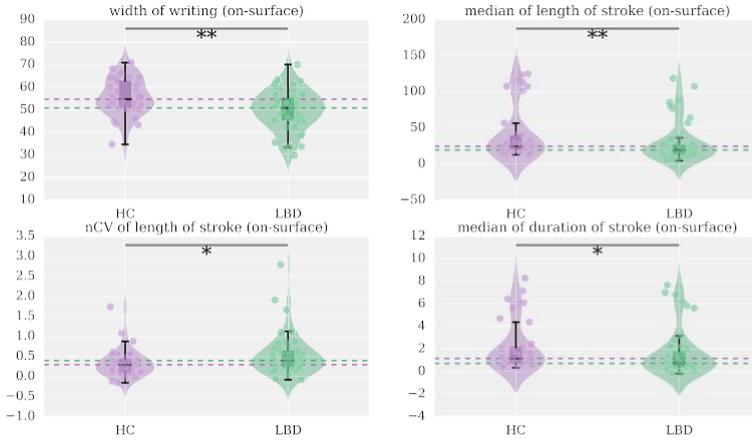

**Fig. 3.** Distribution of the top 4 most discriminating features (pentagons copying test).

**Table 3.** Results of the classification analysis.

| Task | MCC | BACC | SEN | SPE | PRE | F1 | threshold | p |
|---|---|---|---|---|---|---|---|---|
| Spiral | 0.3977 | 0.6848 | 0.8696 | 0.5000 | 0.6349 | 0.7339 | 0.26 | ** |
| Sentence | 0.5271 | 0.7283 | 0.9783 | 0.4783 | 0.6522 | 0.7826 | 0.36 | ** |
| Pentagons | 0.4267 | 0.6848 | 0.9348 | 0.4348 | 0.6232 | 0.7478 | 0.13 | ** |
| All tasks combined | 0.4824 | 0.7391 | 0.8043 | 0.6739 | 0.7115 | 0.7551 | 0.48 | ** |

MCC – Matthew's correlation coefficient; BACC – balanced accuracy; SEN – sensitivity; SPE – specificity; PRE – precision; F1 – F1 score; p – p-values computed by the permutation test (1 000 permutations, ∗ denotes p-value < 0.05 and ∗∗ denotes p-value < 0.01); threshold – fine-tuned decision threshold.

by rigidity. This assumption is supported by the fact that the measure significantly correlates ($\rho = -0.3$, $p < 0.05$) with the overall score of MDS–UPDRS III. Next, the LBD group was not able to keep small variability of loop-to-loop spiral width index, which is in line with findings reported in [31]. We also observed a significant correlation between this feature and the visuospatial ($\rho = -0.4$, $p < 0.01$) and the attention ($\rho = -0.3$, $p < 0.05$) domain of CCB. On the other hand, the LBD group had generally higher values of the spiral precision index than the HC one, which is against our initial assumptions (also the correlation with the attention domain of CCB is surprisingly negative; $\rho = -0.4$, $p < 0.01$). Finally, the last significant correlation with the clinical status was identified in the median of azimuth, which was higher in the LBD group (in addition we observed a negative correlation with the visuospatial domain of CCB; $\rho = -0.4$, $p < 0.05$).

Regarding the classification analysis, based on the spiral features, we were able to discriminate the LBD and HC groups with 68% balanced accuracy (area under the curve (AUC) = 71%), which is the worst result when compared to other tasks and which supports our previous findings that even though the spiral is considered as a gold standard the sentence copy task accents the manifestations of dysgraphia much better [11].

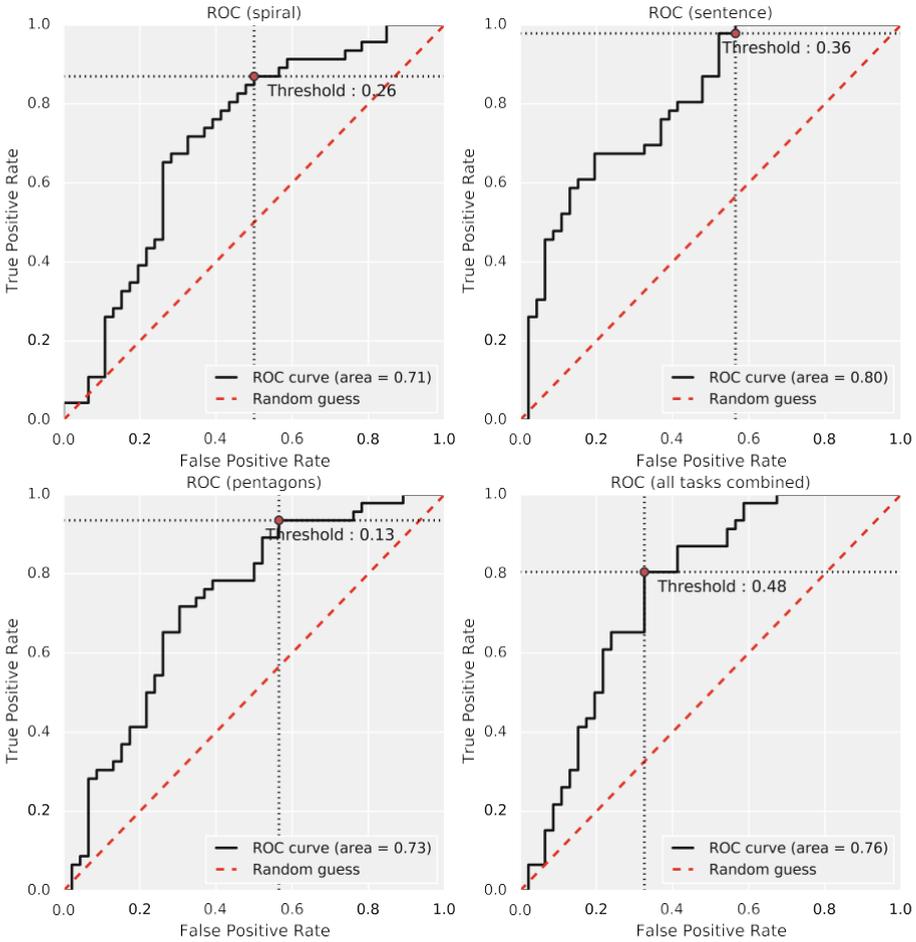

**Fig. 4.** Receiver operating characteristic curves for the trained models.

Regarding the sentence, the most discriminative feature extracted from this task is the number of pen stops (i.e. a pen is in contact with the paper and does not vary its position for at least 30 ms [8]), which was higher in the LBD group. This parameter has been mainly employed in the diagnosis of developmental dysgraphia in children population [27], however, in one study, Danna et al. observed that this measure (but extracted from the spiral) was significantly different between PD patients in the OFF state and HC [8]. Initially, we assumed that the feature could be theoretically linked with cognitive deficits, but we did not observe any significant correlation with the visuospatial, attention, or executive functions domain of CCB. The second most significant feature was the slope of the duration of in-air strokes. The positive correlation coefficient suggests that the LBD subjects were associated with progressing fatigue [1,12,17]. Next, in the LBD group, we observed lower on-surface vertical velocity (this is in line with e.g. [21,35]), but increased on-surface vertical acceleration. This could be probably explained by the slow and less smooth handwriting. In terms of projection, the reason why these deficits dominate in the vertical movement could be explained by the fact that the finger system (which is mainly involved in the vertical movement) is more affected by muscular fatigue than the wrist system (which controls horizontal movement) [20]. The vertical movement requires coordinated movement and finer flexions/extensions of more joints (interphalangeal and metacarpophalangeal), thus it is more complex than ulnar abductions of the wrist [10,34] and could more accent the rigidity and bradykinesia. In addition, this manifestation could be associated with the progressive/consistent vertical micrographia, i.e., progressive/consistent reduction in letter amplitude [33].

In terms of classification, by modelling features extracted from the sentence, we were able to differentiate both groups with 73% balanced accuracy (AUC = 80%). In comparison with the state of the art in supportive LBD or PD diagnosis [9,19,35], it is not a competitive result, but on the other hand, we would like to highlight that we deal with results evaluating diagnosis of LBDs in the prodromal state that has not been targeted by other research teams yet.

Concerning the last (cognitive) task, all the top 5 discriminative features were extracted from the on-surface movement. In our recent article [4] we proved that in-air entropy-based parameters could be used to identify early cognitive deficits in PD without major cognitive impairment and that they correlate with the level of attention. In the current study, these in-air measures were not significant, but on the other hand, their on-surface variants (i.e. median of Shannon entropy calculated from the global/vertical movement) had the p-values of the Mann-Whitney U-test < 0.05, moreover, they significantly correlated with the visuospatial domain of CCB (e.g. $\rho = -0.3$, $p < 0.05$). The top 5 parameters consist of the width of the product, which was smaller in the LBD group. It slightly correlates with the lower median of the length of strokes ($\rho = 0.3$) and lower median of the duration of strokes ($\rho = 0.2$) and probably means that the subjects in the LBD group made the overlapped pentagons smaller. In addition, since the non-parametric coefficient of variation of the length of strokes was higher, we assume that the LBD subjects were not able to keep a stable length of strokes (nevertheless, based on the scoring published in [24], this is assumed as a very small deviation). Regarding the width, we also observed a negative correlation ($\rho = -0.3$, $p < 0.05$) with the overall score of MDS–UPDRS III.

The classification based on the pentagon copying test provided 68% balanced accuracy (AUC = 0.73%), which is slightly better than in the case of the spiral, but not as high as in the case of the sentence.

And finally, a machine learning model based on the whole set of features (tasks) enabled us to improve the accuracy to 74% (AUC = 76%). This shows that the combination of the graphomotor, handwriting and cognitive deficits can be used to achieve reasonable performance in the prodromal diagnosis of LBDs.

## 5   Conclusion

This study has several limitations. Our dataset has a small sample size and the HC and LBD groups are imbalanced, therefore to get better results in terms of their generalisation, a bigger database must be analysed. Next, due to the small sample size, we fused subjects with a high risk of developing PD or MCI-LB into one LBD group. Nevertheless, subjects with MCI-LB in its prodromal stage are associated mainly with cognitive (executive or visuospatial) decline, while subjects with prodromal PD experience mainly motor deficits. In other words, we suppose that further stratification of these participants into two groups could increase the classification accuracy (we hypothesise that MCI-LB would be more pronounced in the pentagon copying task and PD in the handwriting one). Finally, although we tried a correction of multiple comparisons during the statistical analysis, almost no significant features appeared after this adjustment. To sum up, concerning the limitations mentioned above, the study should be considered as a pilot one.

In conclusion, despite the limitations, to the best of our knowledge, it is the first work exploring the impact of computerised analysis of a graphomotor, cognitive, and handwriting task on the prodromal diagnosis of these neurodegenerative disorders. It bridges the knowledge gap in the field of LBDs, and provides baseline results for future studies focusing on the prodromal diagnosis of LBDs via a computerized and objective analysis of graphomotor and handwriting difficulties.